\documentclass[fleqn,final,5p,times,twocolumn,number,sort&compress]{elsarticle}
\usepackage{epsfig}
\usepackage{graphicx}
\usepackage{amsmath}

\usepackage{lineno,hyperref}

\modulolinenumbers[5]

\journal{Journal of Magnetism and Magnetic Materials}

\begin{document}

\begin{frontmatter}
\title{Rotating Zeeman field as a tool for Majorana zero mode
	detection in topological superconducting wire}

\author{Piotr Stefa\'nski}
\address{Institute of Molecular Physics of the Polish Academy of Sciences\\
  ul. Smoluchowskiego 17, 60-179 Pozna\'n, Poland}

\begin{abstract}
 We demonstrate that analysis of the spin polarization of a quantum dot (QD) attached to the topological wire can provide valuable insights into Majorana zero mode (MZM) formation and topological phase transition. Detection is realized by rotation of the  Zeeman field in the wire, while retaining the Zeeman field direction in the dot intact. In the presence of Majorana mode, the effective QD spin polarization at Fermi energy changes significantly when the direction of the Zeeman field in the wire changes from parallel to perpendicular to the wire axis. It can be opposed to the wire in its trivial state, when spin polarization remains practically constant while the magnetic field is rotated. Similar unaltered spin polarization is observed when  QD spin sub-level at Fermi energy mimics MZM. Moreover, the characteristic non-linear dependence of the spin polarization on the magnetic field magnitude at its critical value identifies a topological phase transition in the wire. This feature is observed independently on the coupling strength of the wire to the dot and the angle of the Zeeman field.
\end{abstract}

\begin{keyword}
topological superconductivity\sep Majorana zero modes\sep chiral sub-bands\sep quantum dot
\end{keyword}

\end{frontmatter}

\section{Introduction}
Existence of an exotic fermion, being its own anti-particle, proposed by Ettore Majorana \cite{EttoreMajorana1937}, initiated an extensive search for its possible appearance  in a solid state environment  \cite{Lutchyn2018}. The early theoretical proposals \cite{Lutchyn2010b,Oreg2010a} revealed the mechanism of  formation of  such quasiparticles with Majorana fermion properties, and, as a result, the possibility quantum computation robust to decoherence processes  \cite{DasSarma2015}.

The most suitable scenario for Majorana zero mode (MZM) implementation appeared to be based on  quasi-$1D$ superconductor(Al, NbTiN)-semiconductor(InAs, InSb) heterostructure with strong spin-orbit interaction, subjected to external magnetic field  \cite{Marra2022d}. In February 2025 Microsoft Quantum laboratory announced $Majorana\, 1$, the first $8$-qubit processor chip based on  MZM topological qubits \cite{Majorana1}. The procedure of recognizing true Majorana modes operating in this device was based on the investigation of local and non-local differential conductance of the wire being potentially in topological state. The local conductance  allowed to detect zero bias resonances signaling MZMs. The non-local conductance showed closing-reopening of the superconducting gap, which can indicate  topological phase transition in the wire  \cite{Pikulin2021,Aghaee2022}. 

Very recently an improved $Majorana\, 2$ quantum processor has been presented \cite{Aghaee2026Majorana2Preprint} with an impressive 20 seconds of mean parity lifetime. It is based on a new InAs-Pb heterostructure giving two times larger topological gap as in $Majorana\, 1$.

However, near zero-bias trivial resonances appear to be quite commonly observed \cite{ChenWoods19} due to inhomogeneity of the electrostatic potential in the investigated device.  Moreover, it has been shown that zero bias peaks mimicking Majoranas can be produced by an unintentionally formed quantum dot (QD) in the electrode-wire junction  \cite{Valentini2021} or result from disorder \cite{DasSarmaPan21,ChenGomanko21}. The gap reopening can also be  mimicked by trivial Andreev sub-bands in the wire whose length is not significantly larger than superconducting coherence length \cite{Hess2023}. Thus, the Microsoft announcement initiated a heated debate questioning weather true Majorana modes were observed \cite{Legg2025ParityComment,Legg2025TGPComment,Castelvecchi2026MicrosoftMajorana2NatureNews}.

In the  above context, complementary procedures for MZMs undoubtful screening are still required. 

In this work we propose a method based on spin polarization analysis of the quantum dot attached to the superconducting wire. The measurement is based on the  manipulation of the Zeeman field direction in the wire. As opposed to the original Majorana fermion with a definite spin, the Majorana modes in a solid state matrix exhibit  rather a certain spin polarization, which, as will be demonstrated, depends on the Zeeman field direction in the wire. 

The evolution of MZM spin polarization originates from the existence of  Majorana Kramers pairs, predicted in time-reversal invariant superconductors \cite{Zhang2013}, and formation of chiral sub-bands, caused by strong Rashba interaction in the semiconductor-superconductor wire heterostructure \cite{Alicea2010MajoranaTunableSemiconductor}.

The presented results complement our previous discussion \cite{PStefanski23}  on the detection of the second member of the Kramers pair MZM in the presence of chiral sub-bands. We showed that the optimal conditions for detection of MZM in the second spin sector were a strong QD-wire coupling and tuning of one of the QD's spin-sublevel to Fermi energy. However,  the control of the coupling strength   can be challenging experimentally, and the QD level tuned to Fermi energy can be mistakenly regarded as a Majorana mode. 

There were several proposals of rotating Zeeman field procedure applied in the topological wire for Majorana modes detection. In \cite{Valkov2018} the authors demonstrate electric current switching, caused by MZM spin polarization alteration by magnetic field direction change. The authors of \cite{Bommer2019} investigate experimentally the stability of the induced superconducting gap in InSb nanowires when the direction of Zeeman field changes, in relation to  spin-orbit field. The extended theoretical investigations \cite{Liu2019} complement this work in finding relevant parameters for conductance alteration.

In the present work we demonstrate that even for weak coupling of the dot to the wire and the dot's levels detuned from Fermi energy, detection of the second spin MZM is possible. Indeed, the most preferred configuration is for Fermi energy situated in-between spin-down and spin-up QD sub-levels. 

The proposed detection method is based on a change of the direction of the Zeeman field in the wire and the measurement of the resultant spin polarization in the quantum dot coupled to the wire. We demonstrate, that for a given Majorana mode with a definite chirality, coupled to the dot, the visibility of the MZM particular spin component in the spectral density of the dot is governed by the direction of the Zeeman field in the wire. Specifically, for the Zeeman field in the wire parallel to the Zeeman field in the dot, only one spin component of the Majorana mode is visible, whereas for both the fields mutually perpendicular two spin components of the MZM are manifested. This feature causes remarkable change of the QD spin polarization, in contrast to the case of trivial QD state at Fermi level where polarization is practically constant. Additionally, we show that the characteristic  non-linear dependence of spin polarization  vs. Zeeman field value signals topological phase transition in the wire. We also develop a simple toy model which underpins the essential physics.

\section{Model and computation approach}
The model consists of a quantum dot, created in the junction between a topological wire and a normal electrode. It possesses one active spatial level, which can be shifted electrostatically by the gate voltage. The magnetic field applied along the wire,  in $x$-direction, can be rotated in the $x$-$z$ plane, perpendicular to the effective Rashba field in the wire in $y$-direction. Such a geometry ensures that, independently of the current field direction in the wire, the topological phase can be initiated. The Zeeman field in the dot is set along $z$-direction.

Experimentally it can be challenging to preserve locally the Zeeman field in the dot while rotating the Zeeman field in the wire. However, recently  an  unprecedented control on the spin manipulations in quantum dots by micromagnets in various geometries has been demonstrated \cite{Yoneda15,Stuyck21,Philips22}. 

In the present model we assume that two CoFe magnetic stripes near the quantum dot generate a stable Zeeman field (in $z$-direction), with geometry similar to that utilized in \cite{Jiang25}. Measurement of the spin polarization at Fermi energy is performed by a magnetic STM tip (also magnetized in $z$-direction) hoovering above the dot.  Similar geometry has recently been realized experimentally  \cite{Bagchi24}, for a quantum impurity created in a twin boundaries grain of a MoS${_2}$ single layer  deposited on graphene,  and spin-dependent Kondo effect has been read out. 

Emerging Majorana zero modes can be spotted as zero bias peaks (ZBP) in the differential conductance of the junction.

The quantum dot is described by the Hamiltonian:
\begin{equation}\label{H_QD}
H_{QD}=\sum_{\sigma=\downarrow,\uparrow}\epsilon_{d\sigma}d_{\sigma}^{\dagger}d_{\sigma},
\end{equation}
where the spin sub-levels are split by an external magnetic field: $\epsilon_{d\downarrow/\uparrow}=\epsilon_{d}\mp V_{z}^{QD}$. Zeeman field is $V_{z}^{QD}=(1/2)g\mu_{B}B_{z}$, where $g$ is the effective $g$-factor in the dot and $B_{z}$ is an external magnetic field along $z$ direction. The dot is assumed to be effectively non-interacting due to the large Zeeman splitting provided by two adjacent micromagnets, which mitigates local Coulomb interactions. The semiconducting structure forming the dot  should be preferably of InSb material. It exhibits a large $g$-factor, which would enhance the effect of the magnetic field from micromagnets. In isolated InSb quantum dots $g$-factor of an exceptionally large value $\sim 70$ have been estimated \cite{Nilsson2009InAsNanowires}. In the superconducting environment $g\sim 50$ \cite{Shen2018ParityTransitions}  or $g\sim 20$ \cite{Heedt2021ShadowWall}  has been observed, depending on technology of the device fabrication. In minimal Kitaev chain of InSb- based quantum dots $g\sim 40$ \cite{Dvir2023MinimalKitaevChain}. Generally $g$-factor value depends on the interplay between Rashba and Dresselhaus spin-orbit interactions, the applied gate voltage and the dot radius \cite{DeSousa2003}.

In order to perform tunneling spectroscopy, the dot is coupled to a normal electrode, which is described by the Hamiltonian:
\begin{equation}
H_{lead}+H_{tun}=\sum_{k,\sigma=\downarrow,\uparrow}c_{k\sigma}^{\dagger}c_{k,\sigma}+\sum_{k,\sigma}\left[t_{l}c_{k\sigma}^{\dagger}d_{\sigma}+h.c.\right],
\end{equation}
and introduces the QD level broadening $\Gamma_{l}$.
Finally, the dot is coupled to the first site of the wire:
\begin{equation}
H_{QD-W}=\sum_{\sigma=\downarrow,\uparrow}\left[t_{w} d_{\sigma}^{\dagger}c_{1\sigma}+h.c.\right].
\end{equation}

The wire is set along $x$-direction, subjected to external magnetic field perpendicular to spin-orbit Rashba field. It is described by the Hamiltonian \cite{Stoudenmire2011b,Huang2014a}: $H_{wire}=H_{0}+H_{so}+H_{Z}+H_{X}+H_{sc}$, where:

\begin{eqnarray}\label{Htight}
	H_{0}=\sum_{j=1}^{N}\sum_{\sigma=\downarrow,\uparrow}\epsilon_{w}c_{j\sigma}^{\dagger}c_{j\sigma}
	-t\sum_{j=1,\sigma}^{N-1}\left(c_{j+1\sigma}^{\dagger}c_{j\sigma}+h.c. \right),\\
	H_{so}=\sum_{j=1}^{N-1}\sum_{\sigma,\sigma'}(-i t_{so})c_{j+1\sigma}^{\dagger}\hat{\sigma}^{y}_{\sigma\sigma'}c_{j\sigma'}+h.c.,\\
	H_{Z}=\sum_{j=1}^{N}V_{z}\sin\Theta(c_{j\uparrow}^{\dagger}c_{j\uparrow}-c_{j\downarrow}^{\dagger}c_{j\downarrow}),\\
		H_{X}=\sum_{j=1}^{N}V_{x}\cos\Theta(c_{j\uparrow}^{\dagger}c_{j\downarrow}+c_{j\downarrow}^{\dagger}c_{j\uparrow}),\\
			H_{sc}=\Delta\sum_{j=1}^{N}\left(c_{j\uparrow}^{\dagger}c_{j\downarrow}^{\dagger}+c_{j\downarrow}c_{j\uparrow} \right).
\end{eqnarray}
with $\epsilon_{w}=-(\mu-2t)$.
$H_{0}$ describes the tight-binding part of the Hamiltonian with $t=\hbar^{2}/(2 m^{\star} a^{2})$- the nearest neighbor hopping amplitude between the sites.  $\mu$ denotes chemical potential. $m^{\star}$ is the effective electron mass and $a$ the lattice constant. The operator $c_{j\sigma}^{\dagger}$ ($c_{j\sigma}$) creates (annihilates) an electron of the spin $\sigma$ at the site $j$ of the wire. $H_{so}$ describes the effect of spin-orbit Rashba coupling with $t_{so}=\sqrt{E_{so}t}$, $E_{so}=m^{\star}\alpha^{2}/(2\hbar^{2})$, and $\alpha$- the spin-orbit coupling strength in the wire \cite{Rainis2013}. $H_{sc}$ describes induced superconducting pairing with amplitude $\Delta$, assumed to be real. Finally,  $H_{X}$ and $H_{Z}$ describe Zeeman fields along  $x$ and $z$  direction respectively, governed by the angle $\Theta$ rotation in the $x$-$z$ plane. Zeeman field has the form $V_{x(z)}=(1/2)g\mu_{B}B_{x(z)}$, where $g$ is the effective $g$-factor in the wire, and $B_{x(z)}$ is an external magnetic field applied in a given direction.

In our numerical studies we set the tight-binding hopping to be the energy unit, and relations between other parameters have been chosen to favor topological phase \cite{Stoudenmire2011b,Rainis2013}: $t\gg t_{so}>\Delta$ with $\Delta=0.2$, and $t_{so}=2\Delta$, and chemical potential $\mu=0$. The wire has been assumed to have the length of $N=600$ sites. The topological phase is induced by the increase of the magnetic field, and exists for fields above the critical value $|V|>V^{cr}=\sqrt{\mu^{2}+\Delta^{2}}$  \cite{Lutchyn2010b,Oreg2010a}.

We calculate the Dyson equation matrix for the retarded Green's function of the dot:

\begin{equation}\label{GdMat}
\hat{G}_{d}=\left[\hat{g}_{0}^{-1}-\hat{\Sigma}_{wire} \right]^{-1},
\end{equation}
where $\hat{g}_{0}$ is the bare QD Green's function matrix written in Nambu basis $\Psi=(d_{\downarrow},d_{\uparrow},d_{\uparrow}^{\dagger},d_{\downarrow}^{\dagger})$. It is diagonal with the elements: $\hat{g}_{0}[1,1]=(\omega-\epsilon_{d\downarrow}+i\Gamma_{l})^{-1}$, $\hat{g}_{0}[2,2]=(\omega-\epsilon_{d\uparrow}+i\Gamma_{l})^{-1}$, $\hat{g}_{0}[3,3]=(\omega+\epsilon_{d\uparrow}+i\Gamma_{l})^{-1}$ and $\hat{g}_{0}[4,4]=(\omega+\epsilon_{d\downarrow}+i\Gamma_{l})^{-1}$, where the level broadening  $i\Gamma_{l}$ is due to the coupling to the normal electrode.

The self-energy matrix $\hat{\Sigma}_{wire}$ describes the coupling to the wire. It is expressed by the equation:
\begin{equation}\label{Sig_wire}
\hat{\Sigma}_{wire}=\hat{t}_{w}[\hat{G}]_{1,1}\hat{t}_{w}^{*},
\end{equation}
where $\hat{t}_{w}$ matrix describes the coupling of the dot to the first site of the wire. It has two matrix terms;  diagonal from tight-binding hopping and non-diagonal from Rashba hopping  \cite{Stefanski2021}.

The Green's function matrix $[\hat{G}]_{1,1}$ represents the first site of the wire, and is calculated by recursive summation over the full length of the wire  \cite{Stefanski2021}.

In the following we calculate the spectral density of the dot:
 $\rho_{QD}=-(1/\pi)Im[Tr[\hat{G}_{d}]]$, which can be estimated experimentally from differential conductance measurement.

\section{Chiral sub-bands and evolution of spin under rotating Zeeman field}
In order to describe the spin evolution inside the wire upon rotation of the Zeeman field, we rewrite the wire Hamiltonian in $k$-space. 

After assuming closed periodic boundary conditions we take the expressions of the transformed operators for site $j$: $c_{j,\sigma}=(1/\sqrt{N})\sum_{k}\exp(-i k x_{j})c_{k,\sigma}$ and the representation of the Dirac delta function: $\delta_{k,k'}=(1/N)\sum_{j=1}^{N}\exp[i(k-k')x_{j}]$. The transformed Hamiltonian assumes the form:
\begin{eqnarray}\label{Hwireink1}
	H_{0}=\sum_{k,\sigma}[\epsilon_{w}-2t\cos(k a)]c^{\dagger}_{k\sigma}c_{k\sigma},\\
	H_{R}=2 i t_{so}\sum_{k}(c^{\dagger}_{k\uparrow}c_{k\downarrow}-c^{\dagger}_{k\downarrow}c_{k\uparrow})\sin(k a),\\
H_{Z}=\sum_{k} V_{z}\sin\Theta(c_{k\uparrow}^{\dagger}c_{k\uparrow}-c_{k\downarrow}^{\dagger}c_{k\downarrow}),\\
H_{X}=\sum_{k} V_{x}\cos\Theta(c_{k\uparrow}^{\dagger}c_{k\downarrow}+c_{k\downarrow}^{\dagger}c_{k\uparrow}),\\\label{Hwireink2}
H_{sc}=\Delta\sum_{k}(c_{k\downarrow}c_{-k\uparrow}+c^{\dagger}_{-k\uparrow}c^{\dagger}_{k\downarrow}).
\end{eqnarray}

By introducing the spinor $\hat{c}_{k}=(c_{k\uparrow},c_{k\downarrow})^{T}$, the wire Hamiltonian in $k$-space can be written: 
\begin{eqnarray}\label{Hchiral}
	H=\sum_{k}\hat{c}_{k}^{\dagger}	\begin{bmatrix}
		\tilde{\epsilon}_{w}+V_{z} \sin\Theta & 2 i t_{so}\sin(ka)+V_{x}\cos\Theta \\
	    -2 i t_{so}\sin(ka)+V_{x}\cos\Theta & \tilde{\epsilon}_{w}-V_{z} \sin\Theta 
	\end{bmatrix}\hat{c}_{k},
	\end{eqnarray}
	where  $\tilde{\epsilon}_{w}=\epsilon_{w}-2t\cos(ka)$.
  The $H_{sc}$ term is disregarded here as we are interested in spin evolution under Rashba and Zeeman filed.	
In further considerations it is assumed  $|V_{z}|=|V_{x}|\equiv V$.

Upon diagonalization of the Hamiltonian matrix (\ref{Hchiral}) two chiral sub-bands emerge, with their normalized eigen-vectors:
\begin{eqnarray}\label{Emin}
	\epsilon_{-}=\tilde{\epsilon}_{w}-R, & |\phi_{-}\rangle=\{\frac{-\tilde{t}_{so}+V\cos\Theta}{[2R(R+V\sin\Theta)]^{1/2}},\frac{V\sin\Theta+R}
	{[2R(R+V\sin\Theta)]^{1/2}}\}
\\\label{Eplu}
		\epsilon_{+}=\tilde{\epsilon}_{w}+R, & |\phi_{+}\rangle=\{\frac{-\tilde{t}_{so}+V\cos\Theta}{[2R(R-V\sin\Theta)]^{1/2}},
		\frac{V\sin\Theta-R}{[2R(R-V\sin\Theta)]^{1/2}}\},
\end{eqnarray}
with abbreviations $\tilde{t}_{so}\equiv 2it_{so}\sin(k a)$ and $R=\sqrt{|\tilde{t}_{so}|^{2}+V^{2}}$. The eigenvectors can be written in terms of the eigenvectors of Pauli matrix $\sigma_{z}$  (spin-up and spin-down vectors along $z$-quantization axis):
\begin{eqnarray}\label{alfbet}
	|\phi_{\mp}\rangle=
	\begin{pmatrix}
		\alpha_{\mp} \\
		
		\beta_{\mp}
	\end{pmatrix}=\alpha_{\mp}|\uparrow\rangle+\beta_{\mp}|\downarrow\rangle.
\end{eqnarray}
The dot is subjected to a constant Zeeman field in $z$-direction and can be regarded as a detector of the spin polarization of the particles incoming from the wire. Thus, it is instructive to inspect projections of the eigenvectors of the wire Hamiltonian onto the spin vectors in $z$-quantization. For the lower chiral sub-band, coupled to the dot, we obtain:
\begin{eqnarray}
	|\langle\phi_{-}|\uparrow\rangle|^{2}=|\alpha_{-}|^{2}=\frac{|\tilde{t}_{so}|^{2}+V^{2}\cos^{2}\Theta}{2R(R+V\sin\Theta)},\\
	|\langle\phi_{-}|\downarrow\rangle|^{2}=|\beta_{-}|^{2}=\frac{(R+V\sin\Theta)^{2}}{2R(R+V\sin\Theta)}
.
\end{eqnarray}
In the limit of $k\rightarrow 0$ they become simple expressions:
\begin{eqnarray}
		|\langle\phi_{-}|\uparrow\rangle|^{2}=\frac{1}{2}(1-\sin\Theta),\\
			|\langle\phi_{-}|\downarrow\rangle|^{2}=\frac{1}{2}(1+\sin\Theta).
\end{eqnarray}
For $\Theta=0$ they assume the values $|\langle\phi_{-}|\uparrow\rangle|^{2}=|\langle\phi_{-}|\downarrow\rangle|^{2}=1/2$, and for $\Theta=\pi/2$:   $|\langle\phi_{-}|\downarrow\rangle|^{2}=1$,$|\langle\phi_{-}|\uparrow\rangle|^{2}=0$. This indicates the appearance of MZM modes in both spin sectors for the Zeeman field in the wire along $x$-direction, and MZM is spin down only for the wire Zeeman field along $z$-direction. Furthermore, the effective spin polarization detected by the dot is:
\begin{eqnarray}\label{Polm}
	P_{-}=\frac{|\langle\phi_{-}|\uparrow\rangle|^{2}-|\langle\phi_{-}|\downarrow\rangle|^{2}}
	{|\langle\phi_{-}|\uparrow\rangle|^{2}+|\langle\phi_{-}|\downarrow\rangle|^{2}}=-\sin\Theta
\end{eqnarray}
for $k\rightarrow 0$. Its evolution vs. $\Theta$ angle is similar as obtained from numerical model for topological state of the wire (see Figs.\ref{rhoGammaweak}d and \ref{rhoGammastrong}d). It can be contrasted to the case of trivial state of the wire, when the spin polarization at Fermi energy, detected by the attached dot, is practically independent on the $\Theta$ value (dashed lines in  Figs.\ref{rhoGammaweak}d and \ref{rhoGammastrong}d). When MZMs are absent, the spin polarization is mainly dependent on the Zeeman field in the dot itself and the renormalization of the QD's level caused by the coupling to the wire, which is weakly dependent on $\Theta$. The behavior of the QD polarization for topological wire reveals indirectly  Majorana "leaking" into the dot when topological state of the wire is achieved. 

The same analysis can applied to the upper chiral level $E_{+}$ yielding 
$|\langle\phi_{+}|\uparrow\rangle|^{2}=	|\langle\phi_{-}|\downarrow\rangle|^{2}$, 
$|\langle\phi_{+}|\downarrow\rangle|^{2}=	|\langle\phi_{-}|\uparrow\rangle|^{2}$, and $P_{+}=-P_{-}$.

Discussion of the spin polarization, Eq.~(\ref{Polm}), can be extended to the case of $\Theta=\pi$, when the Zeeman field in the wire is anti-parallel to the $x$-direction. For such arrangement again $|\langle\phi_{-}|\uparrow\rangle|^{2}=|\langle\phi_{-}|\downarrow\rangle|^{2}=1/2$ and MZMs appear in both spin sectors. For $\Theta=(3/2)\pi$ (Zeeman field anti-parallel to $z$-direction) we have $|\langle\phi_{-}|\uparrow\rangle|^{2}=1$ and $|\langle\phi_{-}|\downarrow\rangle|^{2}=0$, indicating that MZM appears in spin-up sector only. 

Finally, let us rewrite the Hamiltonian of the wire, Eqs.~(\ref{Hwireink1})-(\ref{Hwireink2}), in chiral basis, based on Eq.(\ref{alfbet}) (see also \cite{Alicea2010MajoranaTunableSemiconductor}):
\begin{eqnarray}\label{chiralup}
	c_{k\uparrow}=\alpha_{-}c_{k-}+\alpha_{+}c_{k+},\\\label{chiraldown}
		c_{k\downarrow}=\beta_{-}c_{k-}+\beta_{+}c_{k+}.
\end{eqnarray}
It assumes the form $H=H_{0}+\sum_{\gamma=-,+}H_{\gamma}+H_{-+}$:
\begin{eqnarray}
	H_{0}=\sum_{k}(\epsilon_{-}c^{\dagger}_{k-}c_{k-}+\epsilon_{+}c^{\dagger}_{k+}c_{k+}),\\
	H_{\gamma}=\sum_{k}(\Delta_{\gamma}c_{k\gamma}c_{-k\gamma}+
	\Delta_{\gamma}^{\star}c_{-k\gamma}^{\dagger}c_{\gamma}^{\dagger}),\\
	H_{-+}=\sum_{k}(\Delta_{-+}c_{k-}c_{-k+}+\Delta_{-+}^{\star}c_{-k+}^{\dagger}c_{k-}^{\dagger}).
\end{eqnarray}
Superconducting order parameters now describe correlations between chiral particles:
\begin{eqnarray}
	\Delta_{-}=\frac{\Delta}{2R}(\tilde{t}_{so}+V\cos\Theta)=-	\Delta_{+},\\\label{Deltamp}
	\Delta_{-+}=\frac{\Delta V}{RR'}(R\cos\Theta+\tilde{t}_{so}\sin\Theta),
\end{eqnarray}
where $R'=\sqrt{|\tilde{t}_{so}|^{2}+V^{2}\cos^{2}\Theta}$.

Let us analyze the behavior of the order parameters for different angles $\Theta$. Firstly, let us notice that intra-band superconducting order parameter, $\Delta_{-/+}$, has two terms; $p$-wave term with the odd $k$-dependence of $\tilde{t}_{so}$ and $s$-wave term of  strength $\sim\cos\Theta$. The ratio of $|\Delta_{-/+}(\Theta=0)|^2/|\Delta_{-/+}(\Theta=\pi/2)|^2=1+V^2/|\tilde{t}_{so}|^{2}$ shows that the rotation of Zeeman field in the wire from angle $\pi/2$ to $0$ causes an increase of the effective value of the intra-band order parameter by the factor $\sim V^{2}$.

For the inter-band order parameter we obtain $\Delta_{-+}(\Theta=0)=\Delta V/R=-i\Delta_{-+}(\Theta=\pi/2)$, which indicates that for both limiting angles $\Delta_{-+}$ is a pure $s$-wave. Moreover, during the rotation of Zeeman field from $\Theta=0$ to $\Theta=\pi/2$, the inter-order parameter acquires a phase of $\pi/2$. In general, $\Delta_{-+}$ possesses $s$-wave and $p$-wave type contributions with weights $\sim\cos\Theta$ and  $\sim\sin\Theta$, respectively, Eq~(\ref{Deltamp}).

\section{Numerical results}
\subsection{Majorana zero modes in QD's spectral density}
\begin{figure}[h] 
    \centering
	\epsfxsize=\linewidth
	\epsfbox{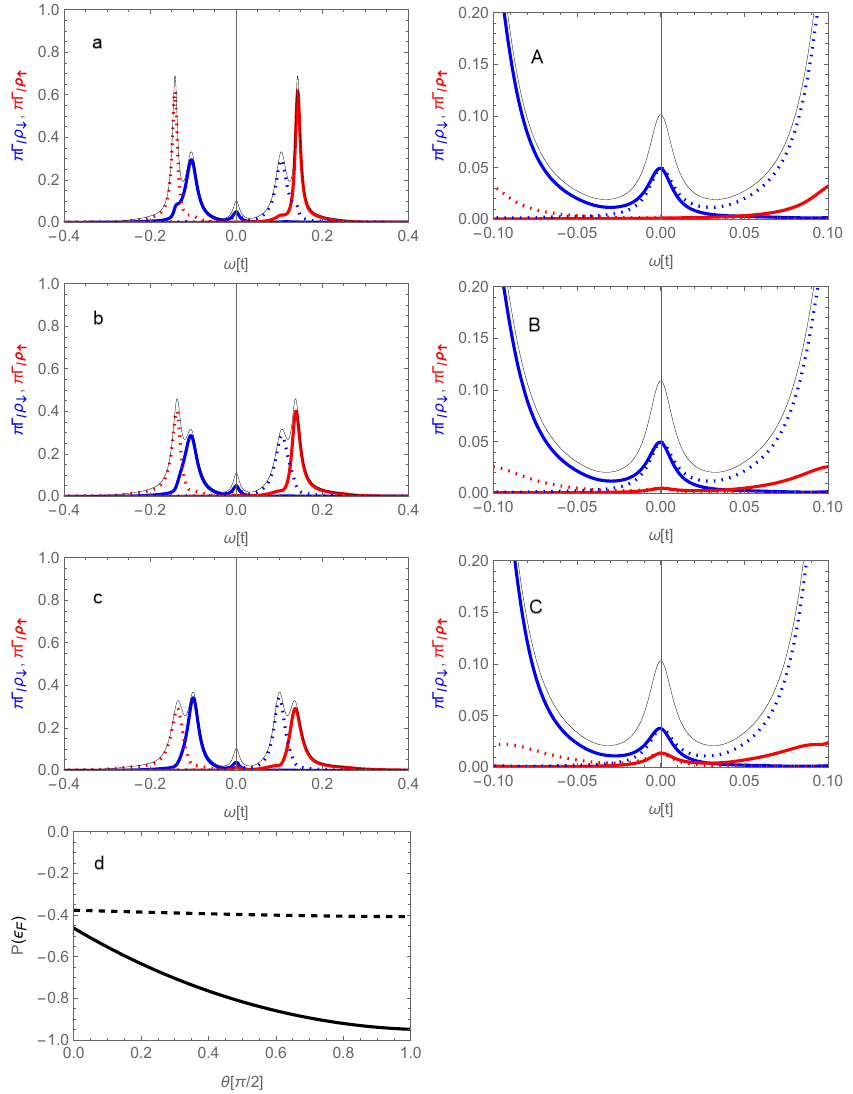}
	\caption{\label{rhoGammaweak}Spectral densities of the dot and its spin polarization at Fermi energy for weak QD-wire coupling, $\Gamma_{s}=0.03$ ($t_{w}=0.20$). Solid (dashed) line - particle (hole) parts, thin solid line - total spectral densities. Blue color is for spin down and red for spin up sector.  Panels a-c (from top to bottom) are for $\Theta=\pi/2,\pi/4$ and $0$. In the right column (panels A-C) are zoomed curves corresponding to the left panels. Panel (d) displays QD spin polarization vs. $\Theta$ angle. Solid curve is for wire in topological state, $V=1.2V_{cr}$, and dashed curve is for the trivial state $V=0.5V_{cr}$.    The remaining input parameters are: $t=1$, $\mu=0$, $\Delta=0.2$, $t_{so}=2\Delta$,  $\epsilon_{d}=0.05$, $V_{z}^{QD}=0.12$, $\Gamma_{l}=0.005$.}
\end{figure}

In Fig.~(\ref{rhoGammaweak}) the results for QD's spectral density and its spin polarization at Fermi energy are presented. The direction of the Zeeman field in the dot is kept intact: along $z$-direction, as the direction of Zeeman field in the wire is changed from  $\Theta=\pi/2$- panel (a), to  $\Theta=\pi/4$- panel (b), and finally to $\Theta=0$- panel (c). In the right column (panels (A)-(C)) corresponding zoomed curves are displayed.  The QD level $\epsilon_{d}$ is adjusted to obtain a renormalized spin sub-level arrangement $\tilde{\epsilon}_{\downarrow}<\epsilon_{F}<\tilde{\epsilon}_{\uparrow}$, and the dot is weakly coupled to the wire, $\Gamma_{s}=0.03$. The peaks of the split QD $\epsilon_{\downarrow}$ and $\epsilon_{\uparrow}$ are localized at $\omega\sim\mp(\epsilon_{\downarrow}+Re\Sigma_{wire}[1,1])$ and  $\omega\sim\mp(\epsilon_{\uparrow}+Re\Sigma_{wire}[2,2])$, respectively, in particle and hole sectors.

For parallel Zeeman fields in the dot and in the wire (panel (a) and (A)), a clear single MZM can be observed at Fermi energy in the spin-down sector; its resonance height is small, due to the fact that the spin down QD level is positioned away from Fermi energy. When Zeeman field becomes tilted (for $\Theta=\pi/4$, panel (b) and (B)) and finally perpendicular (for $\Theta=0$, panel (c) and (C)), the second member of the MZM Kramers pair emerges in the spin-up sector. This feature is clearly observable in the behavior of QD's magnetization at Fermi energy, as presented in the panel (d) (black solid curve). It displays the change of $52\%$ for the present parameters. This is in stark contrast to the case for the wire in the trivial state (dashed curve)- the magnetization remains nearly constant. When one of the dot's spin sub-levels is accidentally at Fermi energy and mimics true MZM, the magnetization is also practically independent on $\Theta$ value and is close to $P(\epsilon_{F})\mp 1$ for $\epsilon_{\downarrow/\uparrow}\simeq\epsilon_{F}$ (see also Toy Model section).

\begin{figure}[h] 
	\centering
	\epsfxsize=\linewidth
	\epsfbox{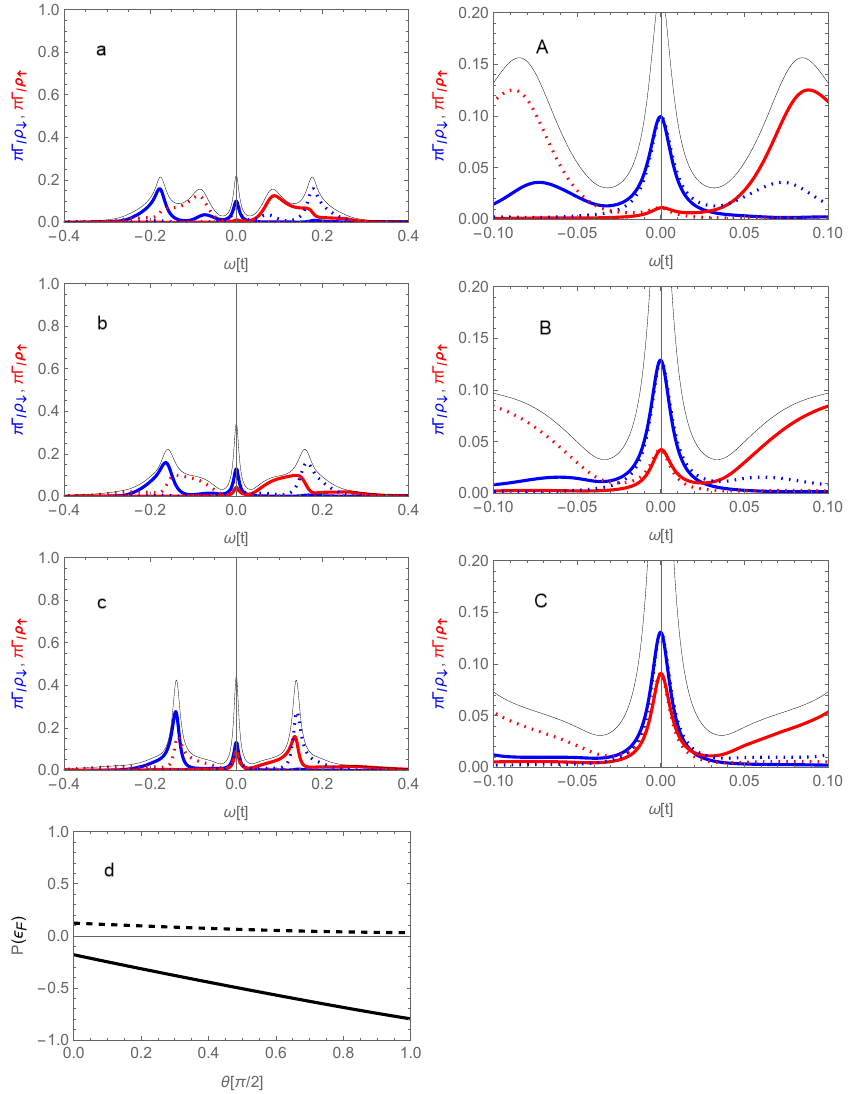}
	\caption{\label{rhoGammastrong}Spectral densities of the dot and its spin polarization at Fermi energy for strong QD-wire coupling,  $\Gamma_{s}=0.15$ ($t_{w}=0.44$) . Solid (dashed) line - particle (hole) parts, thin solid line - total spectral densities. Blue color is for spin down and red for spin up sector. Panels a-c (from top to bottom) correspond to $\Theta_{w}=\pi/2,\pi/4$ and $0$.  In the right column (Panels A-C) are zoomed curves corresponding to the left figures. Panel (d) displays QD spin polarization vs. $\Theta$ angle. Solid curve is for wire in topological state, $V= 1.2V_{cr}$, and  dashed curve is for trivial state $V=0.5V_{cr}$. The remaining input parameters are: $t=1$, $\mu=0$, $\Delta=0.2$, $t_{so}=2\Delta$, $\epsilon_{d}=0.12$, $V_{z}^{QD}=0.12$, $\Gamma_{l}=0.005$.}
\end{figure}

In Fig~(\ref{rhoGammastrong}) spectral density of the dot and its magnetization are presented, for strong QD-wire coupling. The value of quantum dot $\epsilon_{d}$ level is again adjusted by gate voltage to place the Fermi energy level in-between Zeeman split QD sub-levels. Caused by the strong coupling, the manifestation of the second MZM component is more pronounced as the Zeeman field in the wire approaches the direction perpendicular to the QD's Zeeman field. The MZM resonances are better articulated, and the overall change of the QD magnetization reaches $78\%$. Moreover, for the trivial state of the wire,  due to the substantial contribution of the spin-up spectral density at Fermi energy enhanced by stronger coupling, polarization can assume positive values (dashed curve). However, its overall change with $\Theta$ value is practically negligible as compared to curve obtained for the topological state.

\subsection{Indicating topological phase transition}

\begin{figure}[h]
	\centering
	\epsfxsize=\linewidth
	\epsfbox{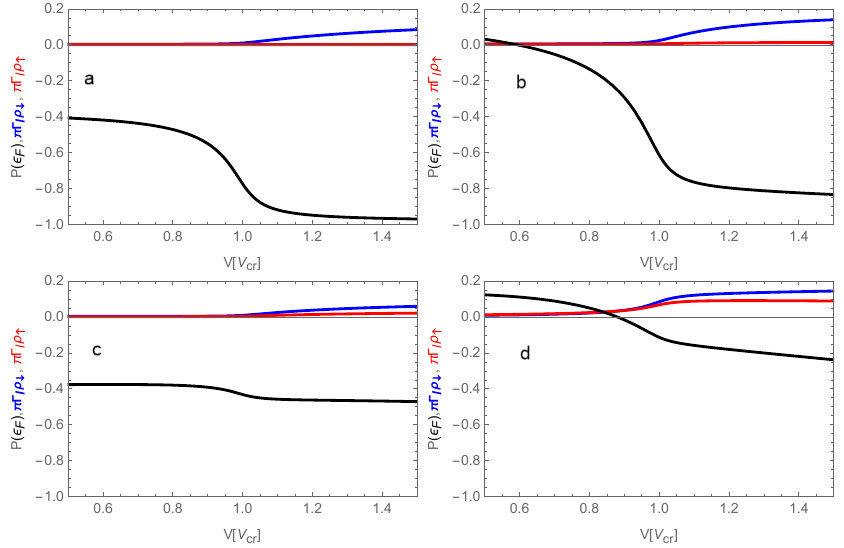}
	\caption{\label{PolTopo}Spectral densities of the dot at Fermi energy and spin polarization vs. Zeeman field in the wire. The red (blue) curves are for spin up (spin down), and black is for magnetization The upper row (panels a and b) is for $\Theta=\pi/2$ and the lower (panels c and d)- for $\Theta=0$. The left column (panels a and c) is for $\Gamma_{s}=0.03$ ($\epsilon_{d}=0.05$) and the right column (panels b and d) is for $\Gamma_{s}=0.15$ ($\epsilon_{d}=0.12$). The remaining input parameters are: $t=1$, $\mu=0$, $\Delta=0.2$, $t_{so}=2\Delta$, $V_{z}^{QD}=0.12$, $\Gamma_{l}=0.005$.}
\end{figure}
Another indication of formation of true Majorana modes  is provided by monitoring  of the QD spin polarization vs. change of the magnetic field value in the wire, depicted in Fig.~(\ref{PolTopo}). A characteristic non-linearity appearing for the same field value, independently of the coupling strength and the angle of Zeeman field, suggests topological transition for this field value. This behavior can be opposed to the case without phase transition, when a linear dependence is rather expected on the change of the Zeeman field. The non-linearity is better exposed for $\Theta=\pi/2$ (panels (a) and (b) of Fig.~(\ref{PolTopo})), when MZM appears in spin-down sector only, but is also clearly visible for $\Theta=0$ (panels (c) and (d)) when MZMs in both spin sectors are present. The strong QD-wire coupling also enhances visibility of the topological transition (in  panels (b) and (d) of Fig.~(\ref{PolTopo})) .

\section{Toy Model}
\begin{figure}[h] 
	\centering
	\epsfxsize=0.8\linewidth
	\epsfbox{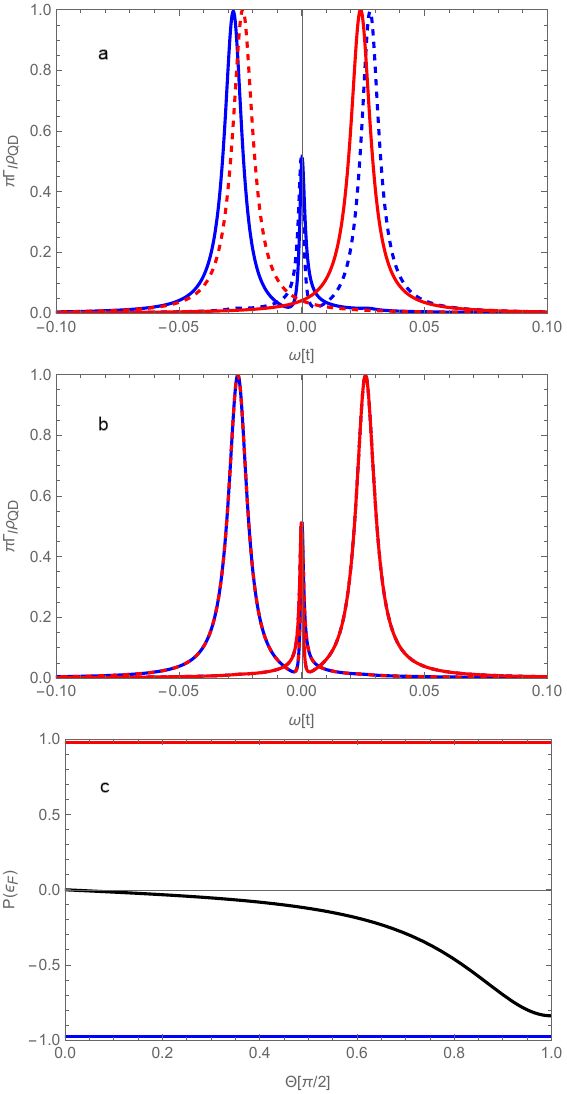}
	\caption{\label{ToyM1}Spectral densities of the dot vs. energy and its spin polarization vs. Zeeman field angle. Panel (a) is for $\Theta=\pi/2$, (b)- for $\Theta=0$. Solid lines-particle, dashed lines-hole components. Blue lines-spin down sector, red lines- spin up sector. The  panel (c) shows spin polarization (black curve); red and blue lines are polarizations for $\epsilon_{\downarrow}=\epsilon_{F}$ and $\epsilon_{\uparrow}=\epsilon_{F}$, respectively. Input parameters are for (a) and (b) are: $\epsilon_{M}=0$, $\epsilon_{d}=0$, $t_{m}=0.01$, $V_{z}^{QD}=0.12$, $\Gamma_{l}=0.005$.}
\end{figure}

\begin{figure} 
	\centering
	\epsfxsize=\linewidth
	\epsfbox{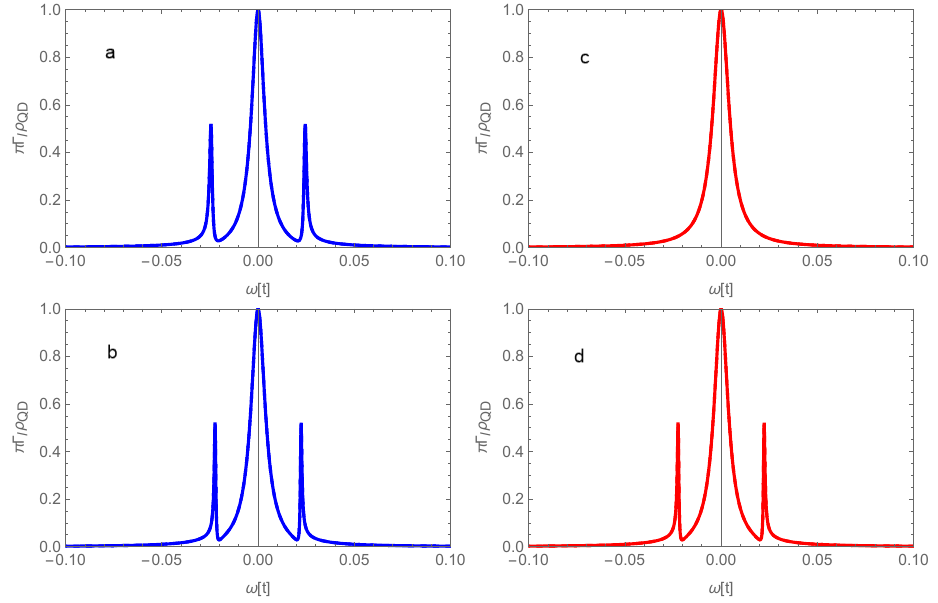}
	\caption{\label{ToyM2}Spectral densities of the dot vs. energy for  $\epsilon_{M}=0.02$. The left column (panels (a) and (b)) is for $\epsilon_{\downarrow}=\epsilon_{F}$ and the right column (panels (c) and (d)) is for $\epsilon_{\uparrow}=\epsilon_{F}$. The  upper row (panels (a) and (c)) is for $\Theta=\pi/2$ the lower (panels (b) and (d)) is for $\Theta=0$. Input parameters are: $\epsilon_{d}=0$, $t_{m}=0.01$, $V_{z}^{QD}=0.12$, $\Gamma_{l}=0.005$.}
\end{figure}

It is instructive to derive a simplified model of the device in order to gain better insight in the physical mechanism present. 
Within the Toy Model, the quantum dot is described similarly as in the numerical model:
\begin{eqnarray}\label{HQDtoy}
	H_{QD}=\sum_{\sigma=\downarrow,\uparrow}\epsilon_{\sigma}d^{\dagger}_{\sigma}d_{\sigma},
\end{eqnarray}
where $\epsilon_{\downarrow/\uparrow}=\epsilon_{d}\mp V_{z}^{QD}$, and the dot's level is broadened by $\Gamma_{l}$ due to the coupling to normal electrode.

It is assumed that each Majorana mode formed  at both ends of the wire belongs to the sub-band of specified chirality. The Hamiltonian of the wire can  then be written as:
\begin{equation}
	H_{w}=i\epsilon_{M}\sum_{j=-,+}\gamma_{1j}\gamma_{2j}.
\end{equation}
It is a natural extension of the low energy "spinless" Hamiltonian \cite{Liu2011DetectingMajoranaQD}, where the spin dependence could be disregarded. 
Then, the Majorana operators can be written in terms of fermionic chiral operators:
\begin{eqnarray}\label{gamma1}
	\gamma_{1j}=(1/\sqrt{2})(c_{j}+c_{j}^{\dagger}),\\\label{gamma2}
		\gamma_{2j}=(-i/\sqrt{2})(c_{j}-c_{j}^{\dagger}).
\end{eqnarray}
In the next step chiral operators are rewritten in terms of fermionic operators with spin index along $z$-direction.  Using Eqs.(\ref{chiralup}) and (\ref{chiraldown}) expressions for chiral operators become:
\begin{eqnarray}\label{cm}
	c_{-}=\beta_{-}c_{\downarrow}+\alpha_{-}c_{\uparrow},\\\label{cp}
		c_{+}=\alpha_{+}c_{\uparrow}+\beta_{+}c_{\downarrow}.
\end{eqnarray}
For $k\rightarrow 0$ the coefficients have simple form:
\begin{eqnarray}
	\alpha_{-}=-\beta_{+}=(1/\sqrt{2})\sqrt{1-\sin\Theta},\\
	\alpha_{+}=\beta_{-}=(1/\sqrt{2})\sqrt{1+\sin\Theta}.
\end{eqnarray}
Taking into account Eqs.~(\ref{cm}) and (\ref{cp})  we get from Eqs.~(\ref{gamma1}) and (\ref{gamma2}): 
\begin{eqnarray}\label{gamma1m}
	\gamma_{1-}=(1/\sqrt{2})[\beta_{-}(c_{\downarrow}+c_{\downarrow}^{\dagger})
	+\alpha_{-}(c_{\uparrow}+c_{\uparrow}^{\dagger})],\\
	\gamma_{2-}=(-i/\sqrt{2})[\beta_{-}(c_{\downarrow}-c_{\downarrow}^{\dagger})
	+\alpha_{-}(c_{\uparrow}-c_{\uparrow}^{\dagger})],
\end{eqnarray}
and
\begin{eqnarray}
	\gamma_{1+}=(1/\sqrt{2})[\alpha_{+}(c_{\uparrow}+c_{\uparrow}^{\dagger})
	+\beta_{+}(c_{\downarrow}+c_{\downarrow}^{\dagger})],\\
	\gamma_{2+}=(-i/\sqrt{2})[\alpha_{+}(c_{\uparrow}-c_{\uparrow}^{\dagger})
	+\beta_{+}(c_{\downarrow}-c_{\downarrow}^{\dagger})].
\end{eqnarray}
Now, the  wire Hamiltonians describing different chiral sub-bands become:
\begin{eqnarray}
	H_{w-}=i\epsilon_{M}\gamma_{1-}\gamma_{2-}=
	(\epsilon_{M}/2)[c_{\uparrow}^{\dagger}c_{\uparrow}+c_{\downarrow}^{\dagger}c_{\downarrow}-\sin\Theta(c_{\uparrow}^{\dagger}c_{\uparrow}-c_{\downarrow}^{\dagger}c_{\downarrow})+\\ \nonumber
	\cos\Theta(c_{\uparrow}^{\dagger}c_{\downarrow}+c_{\downarrow}^{\dagger}c_{\uparrow})-1],
\end{eqnarray}
and
\begin{eqnarray}
	H_{w+}=i\epsilon_{M}\gamma_{1+}\gamma_{2+}=
	(\epsilon_{M}/2)[c_{\uparrow}^{\dagger}c_{\uparrow}+c_{\downarrow}^{\dagger}c_{\downarrow}+
	\sin\Theta(c_{\uparrow}^{\dagger}c_{\uparrow}-c_{\downarrow}^{\dagger}c_{\downarrow})-\\ \nonumber
	\cos\Theta(c_{\uparrow}^{\dagger}c_{\downarrow}+c_{\downarrow}^{\dagger}c_{\uparrow})-1].
\end{eqnarray}
The form of the above equations reveals the underlying physics; the terms $\sim\mp\sin\Theta$ describe the influence of the Zeeman field along the $z$-direction, whereas the terms $\sim\pm\cos\Theta$ correspond to Zeeman field along $x$-direction. Total Hamiltonian $H_{w}=H_{w-}+H_{w+}$ is angle independent:
\begin{eqnarray}\label{Hwtoy}
	H_{w}=\epsilon_{M}\sum_{\sigma=\downarrow,\uparrow}(c_{\sigma}^{\dagger}c_{\sigma}-1).
\end{eqnarray}

The expression of Hamiltonian describing coupling of the QD to the wire is imposed by the direction of the Zeeman field $V_{z}^{QD}$ in the dot in the numerical model and the arrangement of chiral sub-bands in the wire; we assume the coupling of the spin-down dot's sub-level to the Majorana mode of the lower chiral sub-band:
\begin{eqnarray}
	H_{QD-w}=t_{M}(d_{\downarrow}^{\dagger}-d_{\downarrow})\gamma_{1-}.
\end{eqnarray}

In the case of the Zeeman field $V_{z}^{QD}\rightarrow-V_{z}^{QD}$ (and reversal of the Zeeman field in the wire), the corresponding hopping Hamiltonian would be obtained by the exchange: $d_{\downarrow}^{(\dagger)}\rightarrow d_{\uparrow}^{(\dagger)} $, $\gamma_{1,-}\rightarrow\gamma_{1,+}$, $\alpha_{-}\rightarrow\alpha_{+}$, $\beta_{-}\rightarrow\beta_{+}$.

The chiral Majorana operator can be written in terms of its spin components, see Eq.~(\ref{gamma1m}):
\begin{eqnarray}
	\gamma_{1,-}=\beta_{-}\gamma_{1\downarrow}+\alpha_{-}\gamma_{1\uparrow},
\end{eqnarray}
which rescales  coupling of the QD to the wire:
\begin{eqnarray}\label{HQDwtoy}
	H_{QD-w}=t_{M}\beta_{-}(d_{\downarrow}^{\dagger}-d_{\downarrow})\gamma_{1\downarrow}+
	t_{M}\alpha_{-}(d_{\downarrow}^{\dagger}-d_{\downarrow})\gamma_{1\uparrow}.
\end{eqnarray} 

Now, using Eqs.~(\ref{HQDtoy}), (\ref{Hwtoy}) and (\ref{HQDwtoy}) Dyson equation for QD Green's function matrix is constructed:
\begin{eqnarray}
	\hat{G}_{d}(\omega)=\{[\hat{g}_{0}(\omega)]^{-1}-\hat{T}_{M}\hat{F}\hat{T}_{M}^{\star}\}^{-1}.
\end{eqnarray}

The bare Green's function matrix of the dot, $g_{0}$, has the same structure as in the numerical model, see the description below Eq.~(\ref{GdMat}).

The Majorana Green's function matrix, written in the basis $\Psi=(\gamma_{1,\downarrow},\gamma_{1,\uparrow},\gamma_{1,\uparrow},\gamma_{1,\downarrow})$ is of the form
\begin{eqnarray}
\hat{F}=\frac{\omega}{\omega^2-{\epsilon_{M}}^{2}} \left(
\begin{array}{cccc}
	1 & 0 & 0 & 1\\
	0 & 1 & 1 & 0\\
	0 & 1 & 1 & 0\\
	1 & 0 & 0 & 1
\end{array}
\right),
\end{eqnarray}
and the angle dependent hopping matrix is:
\begin{eqnarray}\label{Tmat}
	\hat{T}_{m}=t_{m} \left(
	\begin{array}{cccc}
		\beta_{-} & 0 & 0 & 0\\
		0 & \alpha_{-} & 0 & 0\\
		0 & 0 & -\alpha_{-} & 0\\
		0 & 0 & 0 & -\beta_{-}
	\end{array}
	\right).
\end{eqnarray}

\subsection{Results from Toy Model}

In Fig.~(\ref{ToyM1})  spin-dependent components of the QD spectral density calculated for non-hybridized MZMs, $\epsilon_{M}=0$, are presented. Panel (a), for $\Theta=\pi/2$, displays resonances of the dot level split by Zeeman field with particle and hole components and the MZM resonance at Fermi energy in the spin-down sector only. In panel (b), for $\Theta=0$, additional MZM resonance in spin-up sector appears. The results are in agreement with numerical results, see Figs.~(\ref{rhoGammaweak}a),(\ref{rhoGammaweak}c) and Figs.~(\ref{rhoGammastrong}a),(\ref{rhoGammastrong}c). 

Within the Toy Model this feature can be understood by analyzing the hopping matrix, Eq.~(\ref{Tmat}). For $\Theta=\pi/2$ the $\hat{T}_{m}$ matrix couples the QD and the wire by particle and hole spin-down channel only with the strength of $|t_{m}|$. For $\Theta=0$ all four particle and hole spin channels are open, with the hopping strength $|t_{m}|/\sqrt{2}$.

Panel (c) displays spin polarization at Fermi energy vs. $\Theta$ angle (black curve), which reproduces the tendency observed in numerical results, Fig.~(\ref{rhoGammaweak}d) and Fig.~(\ref{rhoGammastrong}d).

It is also worth to examine the case when a QD's spin sub-level is tuned to Fermi energy and mimics a Majorana resonance. It can be investigated within the Toy Model by setting a finite hybridization,  $\epsilon_{M}=0.02$. In this case MZM resonance disappears. Instead, the Majoranas form an extended fermion manifested as satellite peaks of QD's spin-sublevel. This is shown in Fig.~(\ref{ToyM2}). The left column ((a) and (b)) is for $\epsilon_{\downarrow}=\epsilon_{F}$ and the right column ((c) and (d)) is for  $\epsilon_{\uparrow}=\epsilon_{F}$. For $\Theta=\pi/2$, upper row (panels (a) and (c)), the satellite peaks appear only in spin-down sector, in agreement with only one MZM for $\epsilon_{M}=0$, see Fig.~(\ref{ToyM1}a). For $\Theta=0$, shown in lower row ((b) and (d)), MZM- related satellite peaks are present in the both spin sectors, in agreement with Fig.~(\ref{ToyM1}b). The corresponding spin polarizations vs. $\Theta$ angle, shown in Fig.~(\ref{ToyM1}c), are straight lines, in contrast to the dependence caused by the gradual appearance of the MZM in spin-up sector as the Zeeman field in the wire rotates from $\Theta=\pi/2$ to $\Theta=0$.

The results presented are derived from strictly one-dimensional model. Thus, they are applicable to the experimentally relevant low-density regime, where only one transverse sub-band lies sufficiently close to the Fermi energy and participates in low energy dynamics. I this regime the higher transverse modes in the wire are separated by an energy exceeding the induced superconducting, Zeeman, and spin-orbit energy scales.

\section{Conclusions}
In the present work we presented a alternative  method for true MZM detection, based on the spin dependent analysis of the spectral density of the dot attached to a potentially topological superconducting wire. Due to the strong Rashba interaction in the wire, two chiral sub-bands are formed, and resulting Majorana zero modes share these chiralities. Detection is based on the manipulation of the direction of the Zeeman field in the wire, which is driven from being prependicular to parallel to the wire axis. By this manipulation the second spin component of a given chiral MZM is revealed, which can be observed in spectral density of the dot and the corresponding zero bias differential conductance. We showed that the second spin sector Majorana mode formation remarkably changes spin polarization of the dot, which distinguishes it from the case of accidental QD spin sub-level at Fermi energy level. The detection is robust to the change of the coupling strength between the quantum dot and the wire, and is performed for detuned QD spin sublevels from Fermi energy. 

Additionally, it is shown that the nonlinear dependence of the QD spin polarization on the Zeeman field in the wire signals a topological phase transition at a certain field value, visible independently on the field orientation and coupling strength to the dot. This twofold evidence of the presence of true Majorana modes can facilitate experimental investigations.

\bibliography{libraryPS2}

\end{document}